\documentclass{revtex4}
\usepackage{amsmath}

\input{tcilatex}

\begin{document}

\title[ ]{Specific dynamics for the Domain-Walls in Einstein-Maxwell-Dilaton
theory}
\author{S. Habib Mazharimousavi}
\email{habib.mazhari@emu.edu.tr}
\author{M. Halilsoy}
\email{mustafa.halilsoy@emu.edu.tr}
\affiliation{Department of Physics, Eastern Mediterranean University, G. Magusa, north
Cyprus, Mersin 10, Turkey. }

\begin{abstract}
We consider Einstein-Maxwell-Dilaton (EMD) Lagrangian supplemented by double
Liouville potentials to enrich our system and investigate the resulting
dynamics. The general solution provides us alternative structures induced on
the 3-dimensional domain wall (DW) moving in the 4-dimensional bulk. In
particular, the local maximum in the potential suggests a maximum bounce (or
onset for a contraction phase) of the 3-dimensional
Friedmann-Robertson-Walker (FRW) universe on the DW. Depending on the choice
of parameters we plot various cases of physical interest.
\end{abstract}

\maketitle

\section{Introduction}

The idea that our universe is a brane living in a higher dimensional space
has attracted considerable attention during the recent decade \cite{1}. The
physical properties on such a brane must automatically be induced from the
surrounding space known as bulk through some well - established junction
rules. One such particular example of brane - bulk pair is provided when the
brane has dimension one - less ($d-1$) in a $d-$dimensional bulk known as
domain wall (DW) with the $Z_{2}$ symmetry across. Beside the universe
concept, the DW entails topological defect formations pertaining to the
remnants from the big bang, different vacua of quantum fields and other
applications of vital importance. As emphasized often, one of the most
important problems in modern cosmology is the accelerated expansion of our
currently observed universe, its causes and whether it will do so
indefinitely. Can the answer be obtained from the idea of DWs assumed that
our universe is a DW in a higher dimensional bulk spacetime? In other words,
can we construct an appropriate bulk spacetime so that inflation driven by
matter on the DW is induced from it? It must be admitted that a bulk with
such generality is still missing. Herein we choose our bulk metric to be $4-$%
dimensional ($d=4$) while the DW becomes $3-$dimensional ($d=3$) which may
be considered in this regard as a toy model.

Let us not forget, however, that lower dimensional physics may serve to shed
light on higher dimensions. One justifiable reason for choosing our bulk as $%
d=4$ is that we can provide an exact solution with quite generality to
encompass new dynamics on the DW and serve our purpose well. The action
contains Maxwell and dilaton fields beside gravity whose coupling is
non-minimal \cite{2,3}. With this much rich action, unfortunately, we were
unable to obtain exact solution in bulk dimension $d>4,$ for this reason we
are satisfied with $d=4.$ In addition, we have an extra Liouville type
potential of the dilaton \cite{3} and our Maxwell field is pure magnetic.
Due to the fact that the coupling to dilaton is rather complicated, solution
becomes possible, with a pure magnetic field. Depending on the choice of
integration constants and dilatonic parameter we have obtained a large class
of non-asymptotically magnetic solutions that yield previously known
solutions of its kind in particular limits \cite{4,5,6}. Pure electrically
charged solutions \cite{5}, on the other hand are not expected to overlap
with our magnetic ones. One crucial difference between those and the present
work is in the coupling between dilaton and the magnetic field. Essentially
it is this difference that provides a richer dynamical structure on the DW
without reference to non-physical conditions such as negative mass. It
should also be added that we consider a neutral brane (DW) so that by $Z_{2}$
symmetry, continuity of the vector potential and electromagnetic field
across the brane the Maxwell equations are trivially satisfied. In other
words the surface (DW) Lagrangian doesn't depend on the electromagnetic
field, it depends explicitly only on the dilaton whose boundary conditions
are accounted for in the junction conditions. This amounts to the fact that
the tension / energy density on the brane are only those induced from the
energy and dilaton in the bulk. The effect of the Maxwell fields manifests
itself on the brane through the magnetic charge.

Our DW universe is a $3-$dimensional Friedmann - Robertson - Walker (FRW)
universe with a single metric function (=the radius of the universe)
depending on its proper time \cite{7}. The boundary conditions that connect
bulk to the DW are provided by the Darmoise - Israel conditions \cite{8} apt
for the problem. These conditions determine the energy - momentum on the DW
universe together with its dynamics. An interesting aspect of the solution
that we present in this paper is that on the DW our FRW universe attains
both lower and upper bounces. This particular point provides our main
motivation for this study. It happens that the occurrence of the double
Liouville type potentials in our model made this possible for us. From the
boundary conditions the induced potential on the DW the radius $a\left( \tau
\right) $ of the DW universe satisfies an equation of a particle with zero
total energy (i.e. $\left( \frac{da}{d\tau }\right) ^{2}+U\left( a\right) =0$%
). Such a motion is physical only provided the potential $U(a)$ satisfies $%
U(a)<0$, for all $a(\tau )$. This is investigated thoroughly and we observed
that our DW universe admits both minimum and maximum bounces. Among other
things absence of a DW universe (i.e. $U>0$) is also a possibility.

In Section II we present our model Lagrangian and obtain a general class of
solutions to it. In Section III we analyze the induced potential on the DW.
Conclusion makes our last section IV which summarizes our results.

\section{Our model in $d=4$ and its solution .}

Our action of Einstein-Maxwell Dilaton (EMD) gravity is written as \cite%
{4,5,6,9,10} 
\begin{equation}
S=\frac{1}{\kappa ^{2}}\int_{\mathcal{M}}d^{4}x\sqrt{-g}\left( \frac{1}{2}R-%
\frac{1}{2}\partial _{\mu }\phi \partial ^{\mu }\phi -V\left( \phi \right) -%
\frac{1}{2}W\left( \phi \right) \mathcal{F}\right) +\frac{1}{\kappa ^{2}}%
\int_{\Sigma }d^{3}x\sqrt{-h}\left\{ K\right\} +\int_{\Sigma }d^{3}x\sqrt{-h}%
\mathcal{L}_{DW},
\end{equation}%
where $\mathcal{F}=F_{\lambda \sigma }F^{\lambda \sigma }$ is the Maxwell
invariant$,V\left( \phi \right) =V_{1}e^{\beta _{1}\phi }+V_{2}e^{\beta
_{2}\phi },$ $W\left( \phi \right) =\lambda _{1}e^{-2\gamma _{1}\phi
}+\lambda _{2}e^{-2\gamma _{2}\phi }$ and $\phi =\phi \left( r\right) $ is
the dilaton scalar potential. Herein $\gamma _{i},\beta _{i},V_{i}$ and $%
\lambda _{i}$ are some constants to be identified later while $\mathcal{L}%
_{DW}=-\hat{V}\left( \phi \right) =-V_{0}e^{\epsilon \phi }$ ($\epsilon
=const.$) is the induced potential on the DW. $\left\{ K\right\} $ is the
trace of the extrinsic curvature tensor $K_{ij}$ of DW with the induced
metric $h_{ij}$ ($h=$ $\left\vert g_{ij}\right\vert $). (Latin indices run
over the DW coordinates while Greek indices refer to the bulk's coordinates.
Also in the sequel we use units in which $\kappa ^{2}=8\pi G=1$). The $4-$%
dimensional bulk metric is chosen to be 
\begin{equation}
ds^{2}=-f\left( r\right) dt^{2}+\frac{1}{f\left( r\right) }%
dr^{2}+H(r)^{2}d\Omega _{k}^{2},
\end{equation}%
where $f\left( r\right) $ and $H(r)$ are functions to be found and $d\Omega
_{k}^{2}$ is the line element on a $2-$dimensional space of constant
curvature with $k\in \left\{ -1,0,+1\right\} ,$ i.e., 
\begin{equation}
d\Omega _{k}^{2}=\left\{ 
\begin{array}{c}
d\theta ^{2}+\sin ^{2}\theta d\varphi ^{2}, \\ 
d\theta ^{2}+d\varphi ^{2}, \\ 
d\theta ^{2}+\sinh ^{2}\theta d\varphi ^{2},%
\end{array}%
\begin{array}{c}
k=1 \\ 
k=0 \\ 
k=-1%
\end{array}%
\right. .
\end{equation}

The field equations inside the bulk follow from the variational principles as%
\begin{gather}
R_{\mu }^{\nu }=\partial _{\mu }\phi \partial ^{\nu }\phi +V\left( \phi
\right) \delta _{\mu }^{\nu }+W\left( \phi \right) T_{\mu }^{\nu }, \\
\nabla ^{2}\phi =V^{\prime }\left( \phi \right) +\frac{1}{2}W^{\prime
}\left( \phi \right) \mathcal{F}, \\
\left( \prime \equiv \frac{d}{d\phi }\right) ,  \notag
\end{gather}%
in which 
\begin{equation}
T_{\mu }^{\nu }=2F_{\mu \lambda }F^{\ \lambda \nu }-\frac{1}{2}\mathcal{F}%
\delta _{\mu }^{\nu }
\end{equation}%
is the energy momentum tensor of our Maxwell 2-form $\mathbf{F=}\frac{1}{2}%
F_{\mu \nu }dx^{\mu }\wedge dx^{\nu }$. Variation with respect to the gauge
potential $1-$form $\mathbf{A}$ yields the Maxwell equation

\begin{equation}
\mathbf{d}\left( W\left( \phi \right) ^{\star }\mathbf{F}\right) =0,
\end{equation}%
in which $d\left( .\right) $ is the exterior derivative and the hodge star $%
^{\star }$ means duality. As we commented before, $\phi $ is a function of $%
r,$ and $W\left( \phi \right) =\lambda _{1}e^{-2\gamma _{1}\phi }+\lambda
_{2}e^{-2\gamma _{2}\phi }$ is double Liouville function. This means that
although an electric field ansatz that makes the Maxwell equation
complicated enough for an exact solution, a pure magnetic ansatz easily
satisfies the Maxwell equation. Therefore we prefer to use a magnetic
potential $1-$form with magnetic charge $P$ which reads as (This is also
another feature of our work in comparison with Maity's work in Ref. \cite{5}%
, while in the case of Yazadjiev \cite{6} both electric and magnetic fields
are used)%
\begin{equation}
\mathbf{A}=\left\{ 
\begin{array}{c}
-P\cos \theta \ d\varphi , \\ 
P\ \theta \ d\varphi , \\ 
P\cosh \theta \ d\varphi ,%
\end{array}%
\begin{array}{c}
k=1 \\ 
k=0 \\ 
k=-1%
\end{array}%
\right. .
\end{equation}%
The field $2-$form becomes%
\begin{equation}
\mathbf{F}=\left\{ 
\begin{array}{c}
P\sin \theta \ d\theta \wedge \ d\varphi , \\ 
P\ d\theta \wedge d\varphi , \\ 
P\sinh \theta \ d\theta \wedge \ d\varphi ,%
\end{array}%
\begin{array}{c}
k=1 \\ 
k=0 \\ 
k=-1%
\end{array}%
\right. .
\end{equation}%
where the $^{\star }\mathbf{F}$ is substituted into Eq. (7) and it is easily
satisfied which justifies at the same time our choice of a pure magnetic
gauge potential. For the case of $k=1,$ the magnetic charge $p$ is defined as%
\begin{equation}
p=\frac{1}{4\pi }\oint\limits_{S^{2}}\mathbf{F=}P
\end{equation}%
where $S^{2}$ is the two-dimensional sphere while for the cases $k=0,-1$ the
topological charge takes the form $p=\frac{P\omega }{4\pi },$ in which $%
\omega $ is the area of the corresponding $2-$surface. Based on our choice
of the magnetic field, one can show that independent of $k$ the Maxwell
invariant reads%
\begin{equation}
\mathcal{F}=\frac{2P^{2}}{H(r)^{4}}
\end{equation}%
and the energy momentum tensor becomes 
\begin{equation}
T_{\mu }^{\nu }=\frac{1}{2}\mathcal{F\ }\text{diag}\left[ -1,-1,1,1\right] .
\end{equation}%
Upon substituting these into the field equations (4-5) we get the following
equations%
\begin{eqnarray}
\frac{1}{2H}\left[ f^{\prime \prime }H+2f^{\prime }H^{\prime }\right] &=&-V+%
\frac{1}{2}W\mathcal{F}, \\
\frac{1}{2H}\left[ f^{\prime \prime }H+2f^{\prime }H^{\prime }+4fH^{\prime
\prime }\right] &=&-f\phi ^{\prime 2}-V+\frac{1}{2}W\mathcal{F}, \\
\frac{1}{H^{2}}\left[ HH^{\prime }f^{\prime }+fH^{\prime 2}+fHH^{\prime
\prime }\right] -\frac{k}{H^{2}} &=&-V-\frac{1}{2}W\mathcal{F}, \\
\frac{1}{H^{2}}\left( H^{2}f\phi ^{\prime }\right) ^{\prime } &=&\partial
_{\phi }V+\frac{1}{2}\partial _{\phi }W\mathcal{F}.
\end{eqnarray}%
We note that a prime "$^{\prime }$" implies derivative with respect to $r$
and $\partial _{\phi }=\frac{d}{d\phi }.$ Before we write the solution of
the field equations we comment that our interest here is in finding an
exact, non-asymptotically flat solution which partially was found previously 
\cite{10}. Whether similar attempts may give a different class of
asymptotically flat solution will remain as an open problem.

The general solution for the metric function (after setting $\beta
_{1}=\alpha \sqrt{2},\beta _{2}=\frac{\sqrt{2}}{\alpha },\gamma _{1}=-\frac{%
\alpha }{\sqrt{2}}$ and $\gamma _{2}=\frac{\sqrt{2}}{2\alpha })$ is
expressed by%
\begin{equation}
f\left( r\right) =\left\{ 
\begin{array}{lc}
\begin{tabular}{l}
$\left( 1+\alpha ^{2}\right) ^{2}r^{2}\left[ \frac{P^{2}\lambda _{1}\left( 
\frac{r_{0}}{r}\right) ^{\frac{2\left( 2+\alpha ^{2}\right) }{1+\alpha ^{2}}}%
}{A^{4}\left( 1+\alpha ^{2}\right) }+\frac{\left( \frac{P^{2}\lambda _{2}}{%
A^{4}}-V_{2}\right) \left( \frac{r_{0}}{r}\right) ^{\frac{2}{1+\alpha ^{2}}}%
}{\alpha ^{2}\left( 1+\alpha ^{2}\right) }-\frac{V_{1}\left( \frac{r_{0}}{r}%
\right) ^{\frac{2\alpha ^{2}}{1+\alpha ^{2}}}}{\left( 3-\alpha ^{2}\right) }%
-2M\left( \frac{r_{0}}{r}\right) ^{\frac{3+\alpha ^{2}}{1+\alpha ^{2}}}%
\right] $%
\end{tabular}%
, & \alpha ^{2}-3\neq 0 \\ 
r^{2}\sqrt{\frac{r_{0}}{r}}\left[ \frac{4P^{2}\lambda _{1}}{A^{4}}\left( 
\frac{r_{0}}{r}\right) ^{2}+\frac{4}{3}\left( \frac{P^{2}\lambda _{2}}{A^{4}}%
-V_{2}\right) -4V_{1}\left( \frac{r_{0}}{r}\right) \ln \left( \frac{r}{r_{0}}%
\right) -32M\left( \frac{r_{0}}{r}\right) \right] , & \alpha ^{2}-3=0%
\end{array}%
\right. .
\end{equation}%
Other functions read as

\begin{equation}
H=A\left( \frac{r}{r_{0}}\right) ^{\frac{1}{1+\alpha ^{2}}},\phi =\frac{-%
\sqrt{2}\alpha }{1+\alpha ^{2}}\ln \left( \frac{r}{r_{0}}\right)
,V=V_{1}\left( \frac{r_{0}}{r}\right) ^{\frac{2\alpha ^{2}}{1+\alpha ^{2}}%
}+V_{2}\left( \frac{r_{0}}{r}\right) ^{\frac{2}{1+\alpha ^{2}}},W=\lambda
_{1}\left( \frac{r_{0}}{r}\right) ^{\frac{2\alpha ^{2}}{1+\alpha ^{2}}%
}+\lambda _{2}\left( \frac{r_{0}}{r}\right) ^{\frac{-2}{1+\alpha ^{2}}}
\end{equation}%
where the constant $A$ is related to $P$ by the constraint condition 
\begin{equation}
V_{2}\left( 1-\alpha ^{2}\right) A^{4}+k\alpha ^{2}A^{2}-\lambda
_{2}P^{2}\left( 1+\alpha ^{2}\right) =0.
\end{equation}%
We also note that in the solution (17) $M$ and $r_{0}$ are two integration
constants. We shall use this general solution, with $k=1,$ in the following
Section to construct our $2+1-$dimensional DW. Solution (17) has five model
parameters ($\lambda _{1},\lambda _{2},V_{1},V_{2}$ and $\alpha $) and three
free parameters (after considering the constraint (19)) ($P,M$ and $r_{0}$)
which gives us a large class of BH or non-BH solutions. To get closer to the
solution we give some limits which may be useful in future calculation.
First, we consider $0<\alpha ^{2}<1$ and we find the asymptotic behavior of
the master solution as:%
\begin{eqnarray}
\lim_{r\rightarrow \infty }f &\rightarrow &\left( 1+\alpha ^{2}\right)
^{2}r^{2}\left( -\frac{V_{1}}{\left( 3-\alpha ^{2}\right) }\left( \frac{r_{0}%
}{r}\right) ^{\frac{2\alpha ^{2}}{1+\alpha ^{2}}}\right) =sgn\left(
-V_{1}\right) \infty , \\
\lim_{r\rightarrow 0}f &\rightarrow &\left( 1+\alpha ^{2}\right)
^{2}r^{2}\left( \frac{P^{2}\lambda _{1}}{A^{4}\left( 1+\alpha ^{2}\right) }%
\left( \frac{r_{0}}{r}\right) ^{\frac{2\left( 2+\alpha ^{2}\right) }{%
1+\alpha ^{2}}}\right) =sgn\left( \lambda _{1}\right) \infty ,
\end{eqnarray}%
which show that by choosing proper values for $V_{1}$ and $\lambda _{1}$ we
definitely will obtain BH cases, at least with single horizon. Second, we
consider $\alpha ^{2}=1$ which in the extremal limits admits%
\begin{eqnarray}
\lim_{r\rightarrow \infty }f &\rightarrow &2rr_{0}\left[ \frac{P^{2}\lambda
_{2}}{A^{4}}-V_{2}-V_{1}\right] =sgn\left( \frac{P^{2}\lambda _{2}}{A^{4}}%
-V_{2}-V_{1}\right) \infty , \\
\lim_{r\rightarrow 0}f &\rightarrow &2r^{2}\left( \frac{P^{2}\lambda _{1}}{%
A^{4}}\left( \frac{r_{0}}{r}\right) ^{3}\right) =sgn\left( \lambda
_{1}\right) \infty .
\end{eqnarray}%
The third case is the choice of $\alpha ^{2}>1$ which has the limits as%
\begin{eqnarray}
\lim_{r\rightarrow \infty }f &\rightarrow &\frac{\left( 1+\alpha ^{2}\right) 
}{\alpha ^{2}}r^{2}\left( \frac{P^{2}\lambda _{2}}{A^{4}}-V_{2}\right)
\left( \frac{r_{0}}{r}\right) ^{\frac{2}{1+\alpha ^{2}}}=sgn\left( \frac{%
P^{2}\lambda _{2}}{A^{4}}-V_{2}\right) \infty , \\
\lim_{r\rightarrow 0}f &=&sgn\left( \lambda _{1}\right) \infty .
\end{eqnarray}%
Here also in a similar manner by setting proper values for $\lambda
_{1},\lambda _{2}$ and $V_{2}$ we can construct BH with at least one
horizon. (We note that the limit of the metric for $r\rightarrow 0$ in all
cases are the same.) Among interesting cases one may find that the choice $%
\frac{P^{2}\lambda _{2}}{A^{4}}-V_{2}=0$ gives different asymptotic limits
which for $0<\alpha ^{2}<3$ are given by%
\begin{eqnarray}
\lim_{r\rightarrow \infty }f &\rightarrow &-\left( 1+\alpha ^{2}\right)
^{2}r^{2}\left( \frac{V_{1}}{\left( 3-\alpha ^{2}\right) }\left( \frac{r_{0}%
}{r}\right) ^{\frac{2\alpha ^{2}}{1+\alpha ^{2}}}\right) =sgn\left(
-V_{1}\right) \infty , \\
\lim_{r\rightarrow 0}f &=&sgn\left( \lambda _{1}\right) \infty ,
\end{eqnarray}%
while for $3<\alpha ^{2}$%
\begin{eqnarray}
\lim_{r\rightarrow \infty }f &\rightarrow &-2M\left( 1+\alpha ^{2}\right)
^{2}r^{2}\left( \frac{r_{0}}{r}\right) ^{\frac{3+\alpha ^{2}}{1+\alpha ^{2}}%
}=sgn\left( -2M\right) \infty , \\
\lim_{r\rightarrow 0}f &=&sgn\left( \lambda _{1}\right) \infty .
\end{eqnarray}%
In both cases we have the possibility of choosing proper values for the free
parameters to have BH with at least one horizon. It should also be added
that, to construct a DW in these bulk solutions we choose the radius of the
DW always larger than the possible event horizon. This guarantees that the
DW will not face any singularity problem on its domain.

Let us add that, previously reported solutions are recovered when some
constants of the present model are set to zero, e.g. $\lambda _{2}=V_{2}=0$
leads to the type II solution in \cite{4}, $\lambda _{1}=V_{1}=0$ leads to
the type III solution in \cite{4}, etc.

After all these asymptotic consideration we investigate the form of
singularities at $r=0.$ To do so first we set $r_{0}=1$ and obtain the Ricci
Scalar as 
\begin{equation}
R=\left\{ 
\begin{array}{lc}
\begin{tabular}{l}
$\frac{4A^{2}\left( \alpha ^{4}+\alpha ^{2}+1\right) V_{2}-k\left( \alpha
^{4}+1\right) }{A^{2}\left( 1+\alpha ^{2}\right) ^{2}r^{2/\left( 1+\alpha
^{2}\right) }}+\frac{6\left( \alpha ^{2}-2\right) V_{1}}{\left( \alpha
^{2}-3\right) r^{2\alpha ^{2}/\left( 1+\alpha ^{2}\right) }}-\frac{4\alpha
^{2}M}{\left( 1+\alpha ^{2}\right) ^{2}r^{\left( 3+\alpha ^{2}\right)
/\left( 1+\alpha ^{2}\right) }}+\frac{2P^{2}\alpha ^{2}\lambda _{1}}{\left(
1+\alpha ^{2}\right) A^{4}r^{\left( 4+2\alpha ^{2}\right) /\left( 1+\alpha
^{2}\right) }},$%
\end{tabular}
& 
\begin{tabular}{l}
$\alpha ^{2}\neq 3$%
\end{tabular}
\\ 
\frac{26V_{2}A^{2}-5k}{8A^{2}\sqrt{r}}-\frac{6M-32V_{1}+12V_{1}\ln r}{8r%
\sqrt{r}}+\frac{3\lambda _{1}P^{2}}{2r^{2}\sqrt{r}A^{4}}, & 
\begin{tabular}{l}
$\alpha ^{2}=3$%
\end{tabular}%
\end{array}%
\right. .
\end{equation}%
It clearly shows that in any possible case the solution is singular at $r=0$%
, which for BH cases are screened by horizon(s).

\section{Induced potential on the DW and its implications}

The $3-$dimensional DW on the surface $\Sigma $ in a $4-$dimensional bulk $%
\mathcal{M}$ splits the background bulk into the two $4-$dimensional
spacetimes which will be referred to as $\mathcal{M}_{\pm }.$ Here $\pm $ is
assumed with respect to the DW. Let us look at the master solution (17) and
its asymptotic behaviors given by (20) - (29). We set the parameters such
that $\lim_{r\rightarrow \infty }f\left( r\right) =\infty $ and for the case
of BH we choose $r_{h}<r=a$. For the non-BH case we make the choice $0<r=a$.
Upon imposing the constraint 
\begin{equation}
-f\left( a\right) \left( \frac{dt}{d\tau }\right) ^{2}+\frac{1}{f\left(
a\right) }\left( \frac{da}{d\tau }\right) ^{2}=-1
\end{equation}%
with the DW at $r=a\left( \tau \right) ,$ $\tau $ being the proper time with
respect to the wall observer, the DW's line element takes the form 
\begin{equation}
ds_{dw}^{2}=-d\tau ^{2}+a\left( \tau \right) ^{2}d\Omega _{k}^{2}.
\end{equation}%
This is the standard FRW metric in $3-$dimensions whose only degree of
freedom is $a\left( \tau \right) $, the cosmic scale factor. Now, we wish to
employ the general solution for the field equations (13)-(16) under (11) and
metric ansatz (2). We impose now the rules satisfied by the DW as the
boundary of $\mathcal{M}_{\pm }$. These boundary conditions are the
Darmois-Israel conditions which correspond to the Einstein equations on the
wall \cite{8}. These conditions on the DW $\Sigma $ are given by%
\begin{equation}
-\left( \left\langle K_{i}^{j}\right\rangle -\left\langle K\right\rangle
\delta _{i}^{j}\right) =S_{i}^{j},
\end{equation}%
where the surface energy-momentum tensor $S_{ij}$ is given by \cite{4,5,8}%
\begin{equation}
S_{ij}=\frac{1}{\sqrt{-h}}\frac{2\delta }{\delta g^{ij}}\int d^{3}x\sqrt{-h}%
\left( -\hat{V}\left( \phi \right) \right) .
\end{equation}%
Note that a bracket $\left\langle .\right\rangle ,$ in (33) implies a jump
across $\Sigma $. The stress-energy tensor reduces to the form 
\begin{equation}
S_{i}^{j}=-\hat{V}\left( \phi \right) \delta _{i}^{j}.
\end{equation}%
By employing these expressions through (33) and (34) we find the energy
density and surface pressures for generic metric functions $f\left( r\right) 
$ and $H\left( r\right) $ with $r=a\left( \tau \right) $. The results are
given by%
\begin{equation}
\sigma =-S_{\tau }^{\tau }=-4\left( \sqrt{f\left( a\right) +\dot{a}^{2}}%
\frac{H^{\prime }}{H}\right)
\end{equation}

\begin{equation}
S_{\theta }^{\theta }=S_{\varphi }^{\varphi }=p_{\theta }=p_{\varphi
}=\left( \frac{f^{\prime }+2\ddot{a}}{\sqrt{f\left( a\right) +\dot{a}^{2}}}+2%
\sqrt{f\left( a\right) +\dot{a}^{2}}\frac{H^{\prime }}{H}\right) 
\end{equation}%
in which a dot "$\cdot $" and prime "$^{\prime }$" means $\frac{d}{d\tau }$
and $\frac{d}{da},$ respectively. The Einstein-equations on $\Sigma $
accordingly read 
\begin{gather}
\sqrt{f\left( a\right) +\dot{a}^{2}}\frac{H^{\prime }}{H}=-\frac{1}{4}\hat{V}%
\left( \phi \right) , \\
\frac{f^{\prime }+2\ddot{a}}{\sqrt{f\left( a\right) +\dot{a}^{2}}}+2\sqrt{%
f\left( a\right) +\dot{a}^{2}}\frac{H^{\prime }}{H}=-\hat{V}\left( \phi
\right) .
\end{gather}%
We observe first that%
\begin{equation}
\frac{f^{\prime }+2\ddot{a}}{\sqrt{f\left( a\right) +\dot{a}^{2}}}=\frac{2}{%
\dot{a}}\frac{d}{d\tau }\left( \sqrt{f\left( a\right) +\dot{a}^{2}}\right) 
\end{equation}%
which after using (38) it becomes%
\begin{equation}
\frac{f^{\prime }+2\ddot{a}}{\sqrt{f\left( a\right) +\dot{a}^{2}}}=\frac{2}{%
\dot{a}}\frac{d}{d\tau }\left( -\frac{1}{4}\hat{V}\left( \phi \right) \frac{H%
}{H^{\prime }}\right) =\frac{d}{da}\left( -\frac{1}{2}\hat{V}\left( \phi
\right) \frac{H}{H^{\prime }}\right) .
\end{equation}%
Finally as a result with (39) and (38) yields%
\begin{equation}
\frac{d}{da}\left( \hat{V}\left( \phi \right) \frac{H}{H^{\prime }}\right) =%
\hat{V}\left( \phi \right) .
\end{equation}%
This equation admits a simple relation between $H(r)$ and $\hat{V}\left(
\phi \right) $ given by 
\begin{equation}
H^{\prime }(r)=\xi \hat{V}\left( \phi \right) 
\end{equation}%
with $\xi =$ constant. Upon considering this relation, with (38) and (39)
they become equivalent and therefore we consider the equation (38) alone.
Using above with some manipulation we obtain 
\begin{equation}
\dot{a}^{2}+U\left( a\right) =0
\end{equation}%
where%
\begin{equation}
U\left( a\right) =f\left( a\right) -\frac{1}{16}\frac{H^{2}}{\xi ^{2}}.
\end{equation}%
Further, equations (43) implies%
\begin{equation}
\xi =\frac{1}{\alpha ^{2}+1}\frac{A}{V_{0}r_{0}}
\end{equation}%
and 
\begin{equation}
\hat{V}\left( \phi \right) =V_{0}e^{\frac{\alpha }{\sqrt{2}}\phi }\text{ \
,\ (}V_{0}=\text{cons.)}
\end{equation}

One can easily show that the following boundary condition for the dilaton is
satisfied (modulo Eq. (38) and with reference to \cite{4} (Eq. (38)))
automatically due to the Israel junction conditions 
\begin{equation}
\frac{\partial \phi }{\partial H}=-\frac{2}{H}\frac{1}{\hat{V}\left( \phi
\right) }\frac{\partial \hat{V}\left( \phi \right) }{\partial \phi }.
\end{equation}%
At this stage we consider our solution in two different categories: $\alpha
^{2}\neq 1$ and $\alpha ^{2}=1.$ The latter is known also as a linear
dilaton.

\subsection{The solution for $\protect\alpha ^{2}\neq 1$}

In the sequel we consider the wall to be a classical, one-dimensional
particle moving with zero total energy in the effective potential $U\left(
a\right) .$ For this purpose we employ $f\left( a\right) $ and $H\left(
a\right) $ from the solutions (17) and (18) and to see the general behavior
of the potential $U\left( a\right) $ we rewrite it explicitly as%
\begin{gather}
U\left( a\right) = \\
\left\{ 
\begin{array}{lc}
\begin{tabular}{l}
$\left( 1+\alpha ^{2}\right) ^{2}a^{2}\left[ \frac{P^{2}\lambda _{1}\left( 
\frac{r_{0}}{a}\right) ^{\frac{2\left( 2+\alpha ^{2}\right) }{1+\alpha ^{2}}}%
}{A^{4}\left( 1+\alpha ^{2}\right) }+\frac{\left( \frac{P^{2}\lambda _{2}}{%
A^{4}}-V_{2}\right) \left( \frac{r_{0}}{a}\right) ^{\frac{2}{1+\alpha ^{2}}}%
}{\alpha ^{2}\left( 1+\alpha ^{2}\right) }-\left( \frac{V_{1}}{3-\alpha ^{2}}%
+\frac{V_{0}^{2}}{16}\right) \left( \frac{r_{0}}{a}\right) ^{\frac{2\alpha
^{2}}{1+\alpha ^{2}}}-2M\left( \frac{r_{0}}{a}\right) ^{\frac{3+\alpha ^{2}}{%
1+\alpha ^{2}}}\right] ,$%
\end{tabular}
& 
\begin{tabular}{l}
$\alpha ^{2}\neq 3$%
\end{tabular}
\\ 
\begin{tabular}{l}
$a^{2}\sqrt{\frac{r_{0}}{a}}\left[ \frac{4P^{2}\lambda _{1}}{A^{4}}\left( 
\frac{r_{0}}{a}\right) ^{2}+\frac{4}{3}\left( \frac{P^{2}\lambda _{2}}{A^{4}}%
-V_{2}\right) -4V_{1}\left( \frac{r_{0}}{a}\right) \ln \left( \frac{a}{r_{0}}%
\right) -\left( 32M+V_{0}^{2}\right) \left( \frac{r_{0}}{a}\right) \right] $%
\end{tabular}%
, & 
\begin{tabular}{l}
$\alpha ^{2}=3$%
\end{tabular}%
\end{array}%
\right. .  \notag
\end{gather}%
At this point we introduce a new parameter $\gamma =\frac{1-\alpha ^{2}}{%
1+\alpha ^{2}}$ ($-1<\gamma <1$) and proceed to analyze the forgoing
potential. We set also $r_{0}=1$ for convenience so that the metric function
and the potential for $\alpha ^{2}\neq 3$ ($\gamma \neq -\frac{1}{2}$) take
the form 
\begin{equation}
\left\{ 
\begin{array}{c}
f(a)=\omega _{1}a^{1+\gamma }+\omega _{2}a^{1-\gamma }+\omega _{3}a^{-\gamma
}+\omega _{4}a^{-1-\gamma } \\ 
U(a)=\tilde{\omega}_{1}a^{1+\gamma }+\omega _{2}a^{1-\gamma }+\omega
_{3}a^{-\gamma }+\omega _{4}a^{-1-\gamma }%
\end{array}%
\right.
\end{equation}%
in which%
\begin{equation}
\omega _{1}=\frac{-2V_{1}}{\left( 1+2\gamma \right) \left( 1+\gamma \right) }%
,\omega _{2}=\frac{\left( A^{2}-2P^{2}\lambda _{2}\right) }{A^{4}\gamma }%
,\omega _{3}=-\frac{8M}{(1+\gamma )^{2}},\omega _{4}=\frac{2P^{2}\lambda _{1}%
}{A^{4}\left( 1+\gamma \right) },
\end{equation}%
while 
\begin{equation}
\tilde{\omega}_{1}=\omega _{1}-\frac{V_{0}^{2}}{4\left( 1+\gamma \right) ^{2}%
}.
\end{equation}%
Having the new forms of $f$ and $U,$ one observes that in the domain of $a,$
for $U<f,$ on general grounds there are many possibilities to be considered.
We prefer to consider some cases which are not general but interesting
enough analytically.

The first case is to set $\tilde{\omega}_{1}=\omega _{2}=0$ for which the
potential becomes simply%
\begin{equation}
U(a)=a^{-\gamma }\left( \omega _{3}+\frac{\omega _{4}}{a}\right) .
\end{equation}%
Depending on the sign of $\lambda _{1}$, $M$ and $V_{1}$ different cases may
occur which are shown in Fig. 1. This figure also reveals that the only
configuration, in this setting which admit black hole, with bouncing point
are Fig. 1B and 1F. In Fig. 2 we show these cases in a closer form. The
bouncing points are visible and both are at 
\begin{equation}
a=\frac{\lambda _{1}P^{2}\left( 1+\gamma \right) }{4MA^{4}}.
\end{equation}%
The other interesting setting, at which $U(a)$ may admit two bouncing points
is due to $\tilde{\omega}_{1}=0$ while $\omega _{2}\neq 0.$ In this case the
potential becomes%
\begin{equation}
U(a)=\frac{1}{a^{1+\gamma }}\left( \omega _{2}a^{2}+\omega _{3}a+\omega
_{4}\right)
\end{equation}%
which clearly depends on the values of the parameters and are shown in Fig.
3. In this figure we have introduced 
\begin{equation}
\theta =\frac{\left( P^{2}\lambda _{2}-\frac{1}{2}A^{2}\right) }{A^{4}\gamma 
},\text{ \ }\lambda _{1}^{\left( c\right) }=\left\vert \frac{4\theta
A^{8}M^{2}}{\left( 1+\gamma \right) ^{3}Q^{2}}\right\vert
\end{equation}%
and 
\begin{equation}
V_{1}=\frac{-\left( 1+2\gamma \right) V_{0}^{2}}{1+\gamma }.
\end{equation}%
Let us add that upon setting the latter equality, $V_{0}^{2}$ has no
contribution in the general form of $U(a)$ but still it changes the form of
metric function. This is the reason that Fig.s 3B and 3C (also 3D, 3E and
3F) differ, although the settings are the same. In fact in these figures $%
V_{0}^{2}$ is not the same.

\subsection{the Linear dilaton case $\protect\alpha =1$ ($\protect\gamma =0$)%
}

In the case of linear dilaton with $\alpha =1$ ($\gamma =0$), the form of
the solution is given by%
\begin{equation}
f(r)=2r_{0}^{2}\left[ \frac{P^{2}\lambda _{1}}{A^{4}}\left( \frac{r_{0}}{r}%
\right) +\left( \frac{1}{2A^{2}}-V_{2}-V_{1}\right) \left( \frac{r}{r_{0}}%
\right) -4M\right] ,
\end{equation}%
together with 
\begin{equation}
H^{2}=A^{2}\left( \frac{r}{r_{0}}\right) ,\phi =\frac{-\sqrt{2}}{2}\ln
\left( \frac{r}{r_{0}}\right) ,V=\left( V_{1}+V_{2}\right) \left( \frac{r_{0}%
}{r}\right) ,W=\lambda _{1}\left( \frac{r_{0}}{r}\right) +\lambda _{2}\left( 
\frac{r}{r_{0}}\right) .
\end{equation}%
Herein we used the condition given in Eq. (19) which in the case of $k=1$
and $\alpha =1$ states $2P^{2}\lambda _{2}=A^{2}.$ The DW's potential,
therefore, reads%
\begin{equation}
U\left( a\right) =\frac{1}{a}\left[ \left( \frac{1}{A^{2}}-2\left(
V_{2}+V_{1}\right) -\frac{1}{4}V_{0}^{2}\right) a^{2}-8Ma+\frac{%
2P^{2}\lambda _{1}}{A^{4}}\right] ,
\end{equation}%
where $r_{0}$ is set to one. Here also analytically one can observe that if 
\begin{equation}
\frac{P^{2}\lambda _{1}}{8A^{4}}\left( \frac{1}{A^{2}}-2\left(
V_{2}+V_{1}\right) -\frac{1}{4}V_{0}^{2}\right) <M^{2}<\frac{P^{2}\lambda
_{1}}{8A^{4}}\left( \frac{1}{A^{2}}-2\left( V_{2}+V_{1}\right) \right)
\end{equation}%
and 
\begin{equation}
0<\frac{1}{A^{2}}-2\left( V_{2}+V_{1}\right) -\frac{1}{4}V_{0}^{2}\text{, \ }%
0<\lambda _{1},\text{\ }
\end{equation}%
then there would be two (i.e. both maximum and minimum) bouncing points.
Also if 
\begin{equation}
\frac{P^{2}\lambda _{1}}{8A^{4}}\left( \frac{1}{A^{2}}-2\left(
V_{2}+V_{1}\right) -\frac{1}{4}V_{0}^{2}\right) \geq M^{2}
\end{equation}%
with (62) there would be no dynamical DW universe possible (see Fig. 4)
since $U(a)\geq 0$. Of course there would be all other case also possible
which one can easily find from the closed form of the potential $U(a)$.

With the exception of the cases we investigated closely here many other
cases behave similarly but since they should be treated numerically we do
not study them here. For example the case of $\alpha ^{2}=3$ ($\gamma =-%
\frac{1}{2}$) can be studied only numerically which we ignored here.

\section{Conclusion}

For a pure magnetic field in $d=4$ bulk spacetime we obtain a large class of
solutions in EMD theory. This class generalizes all previously known pure
magnetic type solutions which are obtained in particular limits. The element
that brought new extensions is to take double Liouville type coupling with
dilaton in the action. The junction conditions induce potential on our DW
which is adopted to be a $3-$dimensional FRW universe. An investigation and
plot of the intricate, induced potential reveals the possible existence of a
second (maximum) bounce in our model. This implies an oscillatory universe
on the DW between two (minimum and maximum) limits. In order to go beyond
these bounds and give a big crunch / infinite expansion the DW universe must
naturally undergo quantum tunnelling processes. We recall that in the
Einstein - Gauss - Bonnet bulk in $5-$dimensions, DWs didn't have a second
bounce \cite{11}. A possible extension of our model presented in this paper
may be to consider electromagnetic action, study the Maxwell equations and
find the induced charge on the DW \cite{12}. Let us add that the absence of
any DW universe (i.e. for the potential $U(a)>0$) at all is another extreme
possibility. Finally, it would be much desirable to see the present detailed
analysis of this paper extended to higher dimensions. The difficulty
originates from the fact that the general solution (17-18) obtained in $d=4$
doesn't extrapolate to $d>4$ easily.

\textbf{Acknowledgement:} \textit{We are indebted much to the anonymous
referee for constructive criticism and pointing out numerous errors in the
original version of the paper.}

\bigskip

\textbf{Figure Caption:}

\textbf{Fig. 1:} Diverse plots for different sets of parameters $\lambda
_{1},M$ and $\gamma .$ These range from corresponding black hole states (1A,
1B, 1C, 1E, 1F) to non-black hole states (1D, 1G, 1H) in the bulk. The cases
for $U(a)>0$ (1D, 1H) do not allow formation of DW universes at all. The
figures (1B, 1F) show explicit upper bounces whereas in (1A, 1D, 1E and 1G)
the DW universe has infinite extension. It should be also added that cases
such as $f(a)<0$ (i.e. 1E outside the horizon and 1G every where) do not
correspond to stationary spacetimes rather relate to cosmology.

\textbf{Fig. 2:} With specific parameters we magnify two cases that admit
upper bounces induced from the black hole solutions. Such a bounce enforces
the universe to contract anew. Both 2A and 2B have similar behaviour
although the dilaton parameter differs much.

\textbf{Fig. 3:} A variety of possibilities parametrized by $M$, $\theta $
and $\lambda _{1}$ versus $\lambda _{1}^{\left( c\right) }$, (For
definitions, see Eq. (56) in the text). Depending on the sign of these
parameters we display twelve different cases revealing the available
richness of our DW universe. The bouncing from below (3B), and above (3A),
or both (3D) are explicitly shown. Figures 3C, 3J and 3L are clearly non
stationary.

\textbf{Fig. 4: }Various plots for linear dilaton case $\gamma =0$ (or $%
\alpha ^{2}=1$). Fig.s 4A and 4B are both for non-black hole cases and since 
$U(a)>0$ they don't yield dynamic DW universes. Fig. 4C has the double
bounces but doesn't correspond to a black hole. Fig. 4D corresponds to an
extremal black hole which admits an upper bounce.

\bigskip 

\end{document}